\documentclass[pre,superscriptaddress,twocolumn,amsmath,amssymb,floatfix]{revtex4}

\usepackage{graphicx}
\usepackage{color}

\begin{document}

\title{Dynamical complexity in the perception-based network formation model} 
\author{Hang-Hyun Jo}
\email{johanghyun@postech.ac.kr}
\affiliation{BK21plus Physics Division and Department of Physics, Pohang University of Science and Technology, Pohang 37673, Republic of Korea}
\affiliation{Department of Computer Science, Aalto University School of Science, P.O. Box 15400, Finland}
\author{Eunyoung Moon}
\email{E.Moon@liverpool.ac.uk}
\affiliation{University of Liverpool Management School, Liverpool L69 7ZH, United Kingdom}

\date{\today}

\begin{abstract}
    Many link formation mechanisms for the evolution of social networks have been successful to reproduce various empirical findings in social networks. However, they have largely ignored the fact that individuals make decisions on whether to create links to other individuals based on cost and benefit of linking, and the fact that individuals may use perception of the network in their decision making. In this paper, we study the evolution of social networks in terms of perception-based strategic link formation. Here each individual has her own perception of the actual network, and uses it to decide whether to create a link to another individual. An individual with the least perception accuracy can benefit from updating her perception using that of the most accurate individual via a new link. This benefit is compared to the cost of linking in decision making. Once a new link is created, it affects the accuracies of other individuals' perceptions, leading to a further evolution of the actual network. As for initial actual networks, we consider both homogeneous and heterogeneous cases. The homogeneous initial actual network is modeled by Erd\H{o}s-R\'enyi (ER) random networks, while we take a star network for the heterogeneous case. In any cases, individual perceptions of the actual network are modeled by ER random networks with controllable linking probability. Then the stable link density of the actual network is found to show discontinuous transitions or jumps according to the cost of linking. As the number of jumps is the consequence of the dynamical complexity, we discuss the effect of initial conditions on the number of jumps to find that the dynamical complexity strongly depends on how much individuals initially overestimate or underestimate the link density of the actual network. For the heterogeneous case, the role of the highly connected individual as an information spreader is also discussed.
\end{abstract}


\maketitle

\section{Introduction}

Understanding the structure and dynamics of social networks is crucial to investigate social phenomena that emerge from interaction between human individuals~\cite{Albert2001Statistical, Borgatti2009Network, Goyal2009Connections}. Recent empirical analyses of large-scale datasets on social networks have revealed several common features or stylized facts~\cite{Murase2015Modeling}, such as broad distributions of network quantities~\cite{Onnela2007Analysis}, existence of communities~\cite{Onnela2007Structure, Ahn2010Link, Fortunato2010Community}, and assortative mixing~\cite{Newman2002Assortative}. How and why social networks take such particular forms has been investigated in terms of link formation and/or deletion mechanisms, such as preferential attachment~\cite{Barabasi1999Emergence, Jeong2003Measuring}, triadic and focal closures~\cite{Kossinets2006Empirical, Kumpula2007Emergence, Jo2011Emergence, Murase2014Multilayer}, and link aging~\cite{Murase2015Modeling}. Although these link formation mechanisms have been successful to reproduce various empirical findings in social networks, they have largely ignored the fact that individuals make decisions on whether to create links to other individuals based on cost and benefit of linking. Thus, in this paper, we study strategic link formation mechanisms by considering cost and benefit of linking~\cite{Goyal2009Connections}.

We also observe that the individual perception of the reality could significantly affect the individual behavior. Individuals embedded in a social network may have perception of the network. For example, Milgram conducted the small-world experiment to measure the distance between individuals in the social network~\cite{Milgram1967Smallworld}. In the experiment, each participant was asked to deliver a packet to the target person if he/she knows the target, otherwise to his/her acquaintance who is most likely to know the target. For the latter, the participants may use their own perceptions of the network. Precisely, the individual perception of the network can be straightforwardly represented by an adjacency matrix of the same dimension as that of the network, in alignment with Krackhardt's cognitive social structures~\cite{Krackhardt1987Cognitive}. Using this representation of perception, we can study how one's perception of the network could affect its decision making on whether to create a link to another individual.

We remark that compared to our modeling of individual perception of the network, one can find a number of relatively simple models for human perception or opinion, e.g., in opinion dynamics~\cite{Castellano2009Statistical, Sen2014Sociophysics}. Here a perception or opinion of an individual has been mostly modeled by a spin having several choices or a low-dimensional vector. However, these approaches are often too simplified to properly represent human decision making, despite their successful applications. 

In order to understand the evolution of social networks in terms of perception-based strategic link formation, we study the perception-based network formation model. This model was originally proposed in our previous work~\cite{Jo2015Coevolution}, where we mainly obtained analytic results of the model under restricted conditions. Some of analytic results will be presented in this paper whenever necessary. In this paper we study the model using numerical simulations for more general situations. In the model, each individual has her own perception of the actual network, and uses it to decide whether to create a link to another individual. The perception may not necessarily coincide with the actual network. Then, an individual with the least perception accuracy tries to benefit from updating her perception using that of the most accurate individual via a new link. This benefit is compared with the cost of linking for deciding whether to create a link. Once a new link is created, it affects how accurately other individuals perceive the network, hence their behavior. This might lead to further evolution of the actual network. In this sense, our model can be considered a coevolutionary or adaptive network~\cite{Gross2008Adaptive}. As for initial actual networks, we consider both homogeneous and heterogeneous cases. The homogeneous initial actual network is modeled by Erd\H{o}s-R\'enyi (ER) random networks, while we take a star network for the heterogeneous case. In any cases, individual perceptions of the actual network are modeled by ER random networks with controllable linking probability.

We introduce our model in Section~\ref{sect:model}. In Sections~\ref{sect:homo} and~\ref{sect:hetero} for homogeneous and heterogeneous cases, respectively, we study the dynamics of the model to find that the link density of a stationary network shows discontinuous transitions or jumps according to the cost of linking. Here the structure of stable link density reflects how complex the dynamics of the model is, which we will call a dynamical complexity. Then we discuss the effect of initial conditions on the dynamical complexity in terms of the number of jumps. Finally we make conclusions in Section~\ref{sect:conclusion}.

\section{Model}\label{sect:model}

Let us consider an undirected communication network of $N$ agents who communicate with each other. For a pair of agents $j$ and $k$, the link state $e_{jk,t}$ has the value of $1$ if $j$ and $k$ are connected in a time step $t$, otherwise $0$. We set $e_{ii,t}=0$ for all $i$. The network $G_t$ is a set of link states for all possible pairs, i.e., $G_t=\{e_{jk,t}\}$. We assume that an individual agent observes its own link states with $N-1$ other agents, while guessing link states between other agents that are unobservable to it. Thus, guessed link states may not coincide with the actual link states. Precisely, an agent $i$'s perception $G_t^i$ of the actual network in time step $t$ is a set of perceived link states for all possible pairs, i.e., $G_t^i=\{e^i_{jk,t}\}$, where $e^i_{jk,t}=1$ if $i$ thinks that $j$ and $k$ are connected, otherwise $e^i_{jk,t}=0$. Since each agent correctly knows link states involving itself, we set $e^i_{ij,t}=e_{ij,t}$ for all $j$. A perception accuracy (or accuracy) of an agent $i$ is defined as the aggregated quantity of correct link states in $G^i_t$: 
\begin{equation}
    \rho^i_t= \frac{1}{M}\sum_{j<k}\delta(e_{jk,t}, e^i_{jk,t}),
\end{equation}
where $M=\frac{N(N-1)}{2}$ denotes the maximal number of possible pairs, and $\delta$ is the Kronecker delta function. Due to the fact that $e^i_{ij,t}=e_{ij,t}$ for all $j$, we have
\begin{equation}
  \frac{2}{N}\leq \rho^i_t\leq 1.
\end{equation}

We assume that agents perceiving the network more accurately may use it better, leading to a higher value of a network utility. As agents can observe how well other agents use the network, they can infer the accuracy of other agents. Hence, although the perception, $G_t^i$, is private so that it can be shared only via a link, the accuracy of each agent, $\rho_t^i$, is observable by all other agents. We define a network utility (or utility) of an agent $i$ as an increasing function of the accuracy of $i$, precisely as follows:
\begin{equation}
  u(\rho_t^i)=\rho_t^i,
\end{equation}
where we assume a linearly increasing function for simplicity. As long as the utility is increasing with the accuracy, its functional form is irrelevant to the conclusion. Note that the utility is a function of the accuracy not of perception, implying that the utility of an agent $i$ does not indicate which link states in $G_t^i$ are correct.

The actual network and individual perceptions of it coevolve as follows. In the beginning of each time step $t$, accuracies of all agents, $\{\rho_t^i\}$, are revealed. We first identify a set of agents with the minimum (maximum) accuracy, denoted by $L_t$ ($H_t$). As agents with the minimum accuracy would have the strongest need for improving their perceptions, they try to update their perceptions using those of agents with the maximum accuracy. Let us consider one agent in $L_t$, say $l_t$. In case when there are agents in $H_t$ who are connected to $l_t$, one of them, say $h_t$, is randomly chosen, and $l_t$ does not need to create a new link to $h_t$ for updating its perception. Then, $l_t$ updates its perception using that of $h_t$ as follows:
\begin{equation}
  \label{eq:update}
  e^{l_t}_{jk,t+1}=\left\{\begin{tabular}{ll} 
      $e^{l_t}_{jk,t}$ & if $j=l_t$ or $k=l_t$\\
    $e^{h_t}_{jk,t}$ & otherwise.
\end{tabular}\right.
\end{equation}
Since the accuracy $\rho_t^{h_t}$ does not indicate which link states in $G_t^{h_t}$ are correct, it is best for $l_t$ to replace all of its link states by those of $h_t$ except for link states involving $l_t$ itself. The above perception update will in the next time step lead to 
\begin{equation}
  \label{eq:updatedAccuracy}
  \rho_{t+1}^{l_t}=\frac{2}{N}+\left(1-\frac{2}{N}\right)\rho_t^{h_t}.
\end{equation}
In case when there are no agents in $H_t$ who are connected to $l_t$, $h_t$ is a randomly chosen agent in $H_t$. In this case, $l_t$ needs to create a new link to $h_t$ for a perception update, which is costly. We assume that the cost $c$ of creating a new link is imposed only on $l_t$, and that the maintenance of created links is costless. Then, a new link will be created only when the expected improvement in utility by the perception update via the new link exceeds the cost of linking: 
\begin{equation}
  u(\rho_{t+1}^{l_t})-u(\rho_t^{l_t})>c,
\end{equation}
equivalently,
\begin{equation}
  \frac{2}{N}+\left(1-\frac{2}{N}\right)\rho_t^{h_t}-\rho_t^{l_t}>c,
\end{equation}
where we have used Eq.~(\ref{eq:updatedAccuracy}) because the perception will be updated in the same way as Eq.~(\ref{eq:update}). The left hand side of the above equation is denoted by $c_t$, i.e.,
\begin{equation}
  c_t\equiv \frac{2}{N}+\left(1-\frac{2}{N}\right)\rho_t^{h_t}-\rho_t^{l_t},
  \label{eq:c_t}
\end{equation}
so that the condition for link creation reads $c_t>c$. If $c_t>c$, $l_t$ updates its perception using $h_t$'s perception via a new link between $l_t$ and $h_t$, then the actual network and perceptions by $l_t$ and $h_t$ are updated with $e_{l_th_t,t}= e^{l_t}_{l_th_t,t}= e^{h_t}_{l_th_t,t}=1$. Otherwise, if $c_t\leq c$, nothing happens. The cases are summarized as follows:
\begin{enumerate}
  \item If $e_{l_th_t,t}=1$, $l_t$ updates its perception using $h_t$'s.
  \item If $e_{l_th_t,t}=0$ and $c_t>c$, $l_t$ updates its perception using $h_t$'s via a new link between $l_t$ and $h_t$.
  \item If $e_{l_th_t,t}=0$ and $c_t\leq c$, nothing happens.
\end{enumerate}
Then, the time step $t$ ends. Note that for each new link, say $ij$, the accuracy of an agent other than $i$ and $j$ must either increase or decrease by $\frac{1}{M}$.

Since there can be more than one agent with the minimum accuracy in $L_t$, we consider two rules one by one: (i) Only one agent randomly chosen in $L_t$ has a chance to update its perception, which we call a \emph{single update}. (ii) The chance is given to all agents in $L_t$ simultaneously, which we call a \emph{multiple update}. Thus, the network will evolve faster under the rule of the multiple update than under the rule of the single update. 

As for initial actual networks, we will consider homogeneous and heterogeneous cases. For the homogeneous initial actual network, we use an Erd\H{o}s-R\'enyi (ER) random network with linking probability $q$, where $e_{jk,0}=1$ ($0$) with probability $q$ ($1-q$). If $q=0$, the initial network is empty. For the heterogeneous initial actual network, we use a star network, where one agent, called center, is connected to all other agents. The link density of a star network is $\frac{2}{N}$. In all cases, the initial perception by each agent $i$ is assumed to be an ER random network with linking probability $p$ except for the links involving the agent as following: 
\begin{equation}
    e^i_{jk,0}=\left\{\begin{tabular}{ll}
      $e_{jk,0}$ & if $j=i$ or $k=i$,\\
      $0$ & if $j,k\neq i$, with prob. $1-p$,\\
      $1$ & if $j,k\neq i$, with prob. $p$.
  \end{tabular}\right.
\end{equation}
If the value of $p$ is larger than the link density of the initial actual network, i.e., $q$ or $\frac{2}{N}$, it can be interpreted such that individuals overestimate the number of links in the actual network. We will show that the degree of overestimation strongly affects the dynamical complexity, depicted by the complexity of $c_t$.

\begin{figure}[!t]
  \includegraphics[width=\columnwidth]{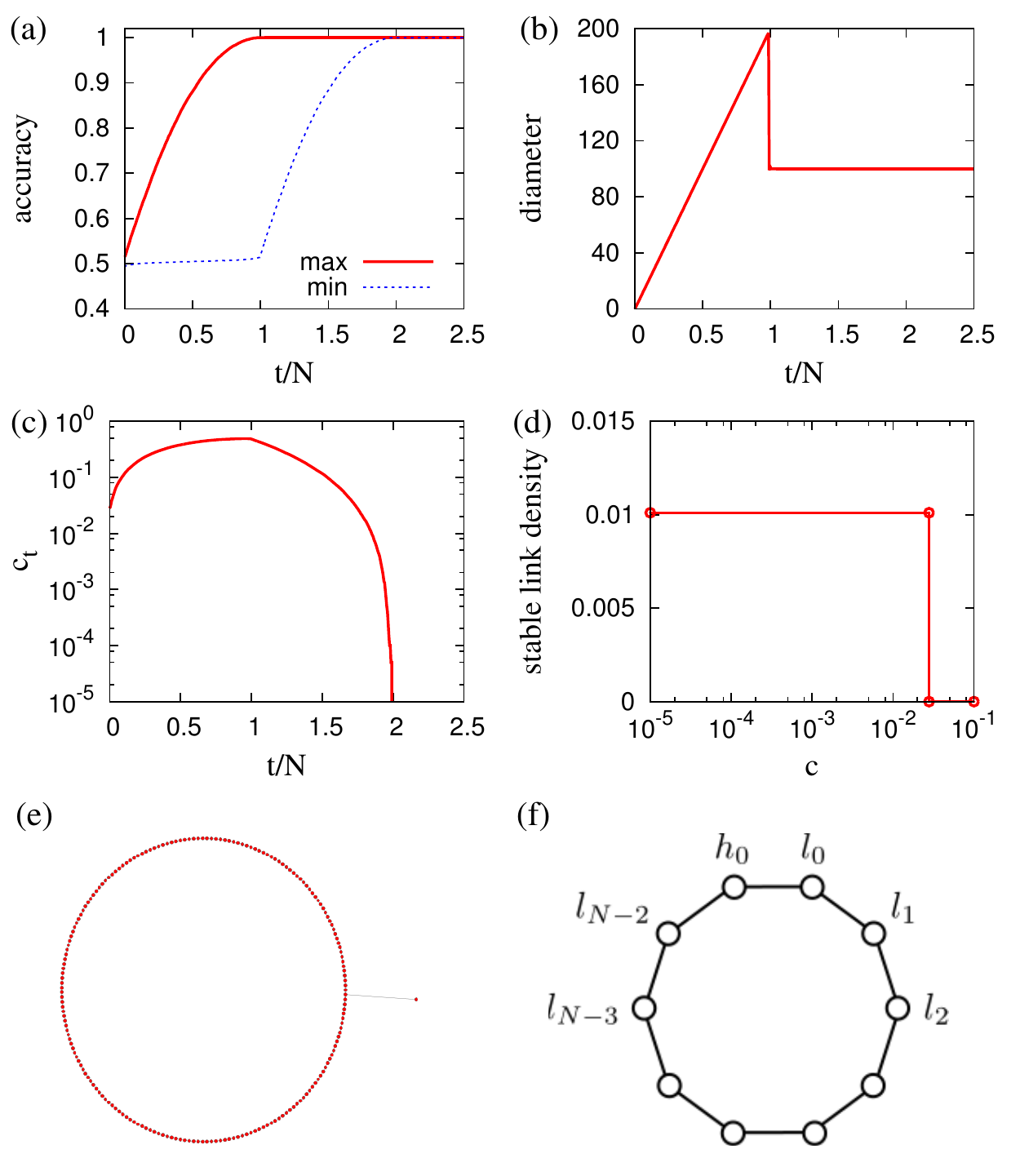}
  \caption{Single update case for an initially empty network for $N=200$ and for $p=0.5$. We present temporal evolutions of (a) maximum and minimum accuracies, (b) the diameter of the actual network, and (c) the value of $c_t$ in Eq.~(\ref{eq:c_t}), with the resulting stable link density as a function of $c$ in (d). The stationary network for $c=0$ is visualized in (e). The schematic diagram for a ring structure of the actual network is depicted in (f).}
    \label{fig:oneUpdate_empty_p0_5}
\end{figure}

\section{Homogeneous initial condition: random networks}\label{sect:homo}

For numerical simulations, we set $N=200$ to take the finite capacity of an individual perception into account, as the maximal number of links in each agent's perception is of the order of $N^2$. We however remark that our results remain the same independent of $N$ as long as $N$ is finite.

\subsection{Single update case}

We consider the single update case in which at most one link can be created in each time step. As a benchmark, we study the case with $q=0$, i.e., an empty network as the initial actual network. It indicates that all agents are strangers to each other in the beginning. In order to describe the early stage of the dynamics in the simplest form, $H_t$ is assumed to have only one agent for $0\leq t<N$. This is indeed the case when $p$ is sufficiently large, e.g., when $p=0.5$, as shown in Fig.~\ref{fig:oneUpdate_empty_p0_5}. It is because the larger $p$ tends to result in the wider range of accuracies, and the smaller number of agents with maximum accuracy. We demonstrate the simplest case of the large $p$ before discussing the effect of smaller $p$. 

Since the actual network stops evolving in a time step $t$ when $c_t\leq c$ and $e_{l_th_t,t}=0$ for every agent $l_t$ in $L_t$, the control parameter $c$ determines when to stop the evolution of the actual network. We will first set $c=0$ to obtain the dynamics of $c_t$ until when all agents have the same accuracy, implying $c_t=0$. Then we will discuss the effect of positive $c$. Given that $c=0$, the first link is created between the most and least accurate agents, $h_0$ and $l_0$. Then the perception of $l_0$ is replaced by that of $h_0$ except for link states involving $l_0$ itself. Since Eq.~(\ref{eq:updatedAccuracy}) implies $\rho^{l_0}_1\geq \rho_0^{h_0}$, $l_0$ is most likely to become the most accurate agent in $t=1$, i.e., $h_1=l_0$. In $t=1$, the least accurate agent $l_1$ creates a link to $l_0$ to become the most accurate agent in $t=2$, i.e., $h_2=l_1$. In general, we find that $h_{t+1}=l_t$, and that the network evolves in a line connecting $h_0$, $l_0$, $l_1$, $\cdots$, $l_{t-1}$ sequentially. Here the correct link states involving $l$s are cumulated along the line network so that the maximum accuracy increases as $t$ increases, see Fig.~\ref{fig:oneUpdate_empty_p0_5}(a). In $t=N-2$, the network is connected by the link between $l_{N-3}$ and $l_{N-2}$, and the accuracy of $l_{N-2}$ becomes $1$. It means that the initially incorrect link states of $h_0$ are completely replaced by correct link states, which are yet carried only by $l_{N-2}$. Then, in $t=N-1$, the least accurate agent $h_0$ creates a link to the most accurate agent $l_{N-2}$, leading to a ring structure in the network, see Fig.~\ref{fig:oneUpdate_empty_p0_5}(f). This is consistent to the numerical observation that the diameter drops from $\approx N$ to $\approx \frac{N}{2}$ as shown in Fig.~\ref{fig:oneUpdate_empty_p0_5}(b). From now on, we denote $N$ agents in the network by $h_0$, $l_0$, $\cdots$, $l_{N-2}$, respectively. 

For $N\leq t<2N$, we assume that $|L_t|=1$ again due to the large $p$. As $h_0$ updated its perception by that of $l_{N-2}$ in $t=N-1$, $h_0$ has fully correct link states of the actual network, i.e., $H_N=\{l_{N-2},h_0\}$, while the least accurate agent is $l_0$. Since $l_0$ and $h_0$ are neighboring, $l_0$ updates its perception by that of $h_0$ without creating a new link. In the next time step, $l_1$ is the least accurate agent and updates its perception by that of $l_0$, and so on. During this process the fully correct link states propagate along the ring without creating any links, then finally in the time step $t=2N-2$, all agents share the fully correct link states of the actual network, leading to $c_{2N-2}=0$. Accordingly, the average link density of perceptions converges to the link density of the actual network.

The temporal evolution of $c_t$ is determined by the maximum and minimum accuracies. As analyzed in~\cite{Jo2015Coevolution}, the behavior of $\rho_t^{h_t}$ is described by
\begin{eqnarray}
\rho_t^{h_t}=\left\{\begin{tabular}{ll}
$1-\frac{(N-t-1)(N-t-2)}{(N-1)(N-2)}\left(1-\rho_0^{h_0}\right)$ & \mbox{if} $t<N-1$,\\
$1$ & \mbox{if} $t\geq N-1$.
\end{tabular}\right.
\end{eqnarray}
Here the maximum accuracy is only a function of $\rho_0^{h_0}$. In contrast, the temporal evolution of the minimum accuracy depends on the initial accuracies of all agents other than $h_0$ for $0\leq t<N$. Since the fluctuation in $\rho_t^{l_t}$ is negligible and $c_t$ is driven by $\rho_t^{h_t}$, $c_t$ overall increases until $t\approx N$. For $N\leq t<2N-1$, as $\rho_t^{h_t}=1$, we have $c_t=1-\rho_t^{l_t}$ with $\rho_t^{l_t}=\rho_{t'}^{h_{t'}}$ for $t'=t-N+1$~\footnote{For example, in $t=N$, the least accurate agent is $l_N=l_0$, whose accuracy in $t=N$ is the same as $\rho_1^{l_0}$. With $h_1=l_0$, we have $\rho_N^{l_N}=\rho_N^{l_0}=\rho_1^{l_0}=\rho_1^{h_1}$. In general, $\rho_t^{l_t}=\rho_{t-N+1}^{h_{t-N+1}}$ for $N\leq t<2N-1$.}. Then $c_t$ finally decreases to $0$, comparable to numerical simulations shown in Fig.~\ref{fig:oneUpdate_empty_p0_5}(c). From the dynamics of $c_t$, one can infer a stationary network structure as a function of $c$. For sufficiently large $c$, i.e., if $c\geq c_0$, even the first link cannot be created, leading to the empty network as a stationary network. If $c<c_0$, the network has a chance to evolve to a ring structure, whose link density is $\frac{2}{N-1}$. Thus, the stable link density as a function of $c$ shows a discontinuous transition or a jump at $c=c_0$, as depicted in Fig.~\ref{fig:oneUpdate_empty_p0_5}(d).

\begin{figure}[!t]
  \includegraphics[width=\columnwidth]{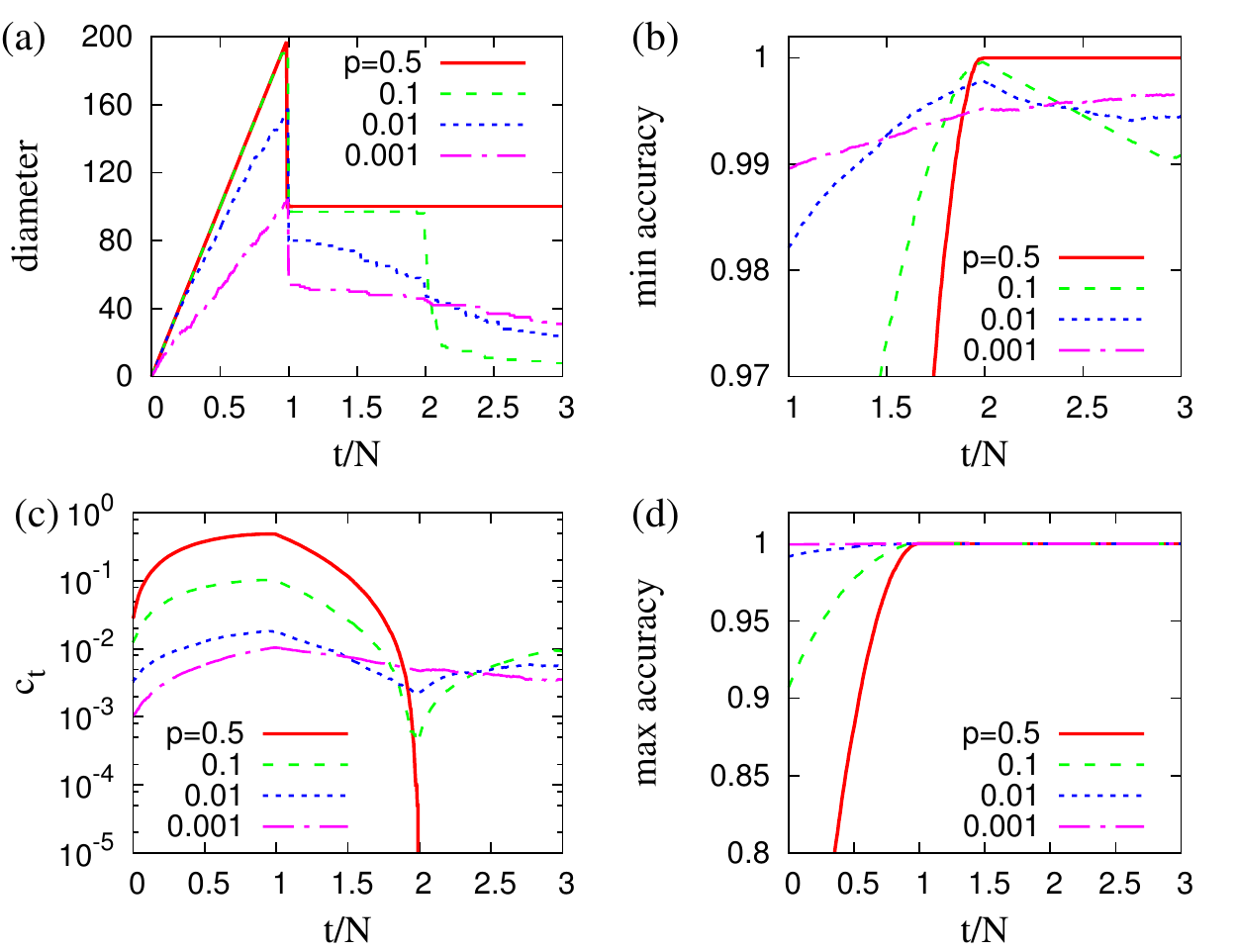}
  \caption{Single update case for an initially empty network for $N=200$ and for various values of $p$. We present temporal evolutions of (a) diameters of actual networks, (b) minimum accuracies, (c) values of $c_t$ in Eq.~(\ref{eq:c_t}), and (d) maximum accuracies.}
    \label{fig:oneUpdate_empty_comparison_p}
\end{figure}

\begin{figure}[!t]
  \includegraphics[width=\columnwidth]{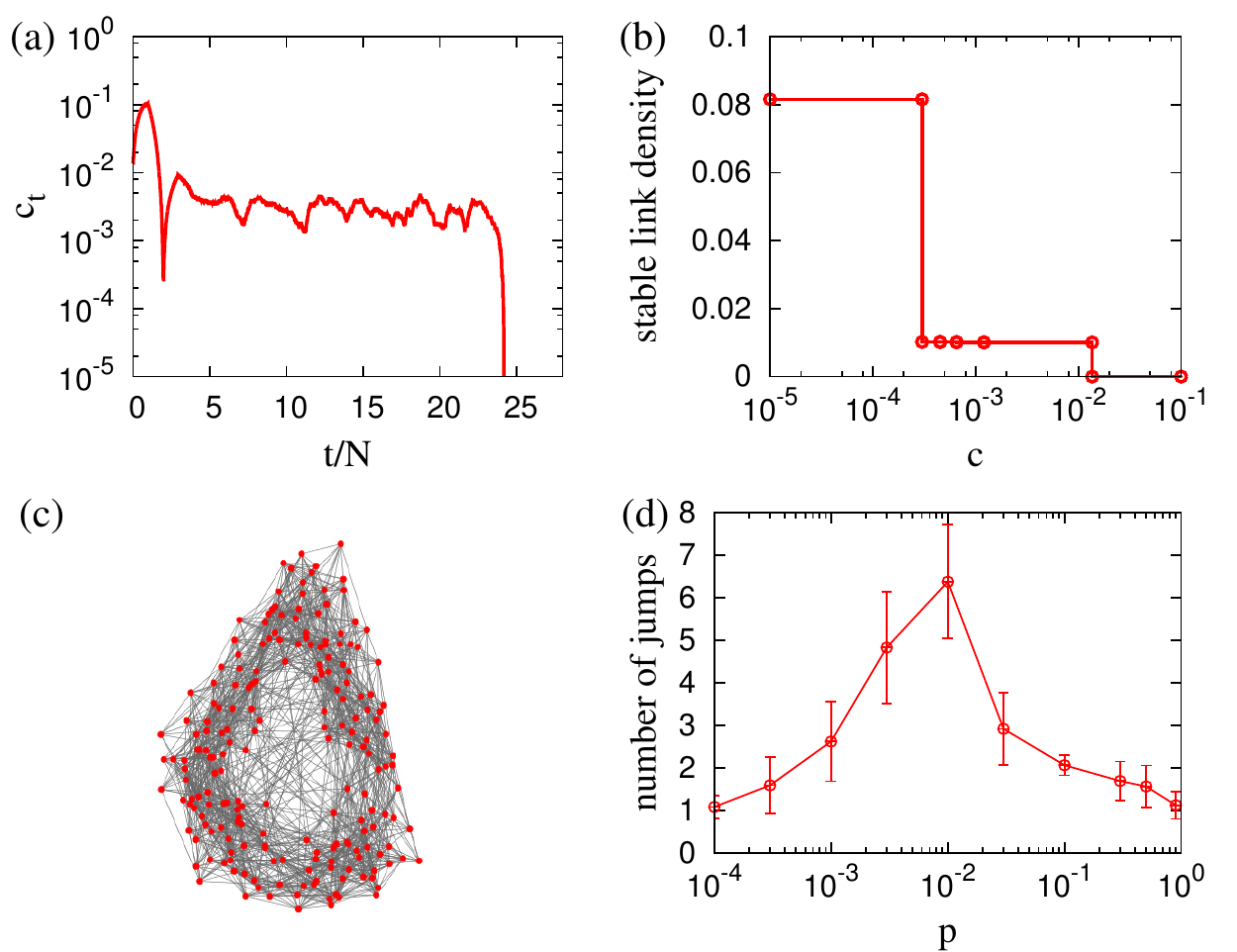}
  \caption{Single update case for an initially empty network for $N=200$ and for $p=0.1$ (a--c). A temporal evolution of $c_t$ in Eq.~(\ref{eq:c_t}) is presented in (a) with the resulting stable link density as a function of $c$ in (b). The stationary network for $c=0$ is visualized in (c). In (d), the average number of jumps of stable link density as a function of $c$ over $100$ runs is plotted with standard deviation for each value of $p$.}
    \label{fig:oneUpdate_empty_p0_1}
\end{figure}

Now we discuss the effect of small values of $p$ on the dynamics. When $p$ is large, the initial accuracies become more distinct from each other, which enables that $|H_t|=1$ for $0\leq t<N$ and that $|L_t|=1$ for $N\leq t<2N$. However, when $p$ is small, i.e., when the initial perceptions are relatively similar to the initial actual network, the multiplicities of $H_t$ and $L_t$ lead to more complicated evolution of the actual network. For $0\leq t <N$, the multiplicity of $H_t$ may lead to a number of branches along the line network before forming a ring structure. Accordingly, the diameter of the network increases more slowly, as shown in Fig.~\ref{fig:oneUpdate_empty_comparison_p}(a). For $N\leq t<2N-1$, the multiplicity of $L_t$ may lead to the formation of loops. For example, consider a segment of the ring network such that $e_{l_k,l_{k+1},t}=e_{l_{k+1},l_{k+2},t}=1$ and $e_{l_k,l_{k+2},t}=0$. When $H_t=\{l_k\}$, $L_t=\{l_{k+1},l_{k+2}\}$, and $l_{k+2}$ is given the chance to update, $l_{k+2}$ will create a link to $l_k$ to form a loop. This new link lowers accuracies of agents whose link state on $l_k l_{k+2}$ was correct. These accuracies could be lowered further because such new link can later provoke more new links. As a result, the minimum accuracy does not reach $1$ in $t\approx 2N$ in Fig.~\ref{fig:oneUpdate_empty_comparison_p}(b). As shown in Fig.~\ref{fig:oneUpdate_empty_comparison_p}(c), $c_t$ does not decay to $0$ in $t\approx 2N$ with the fact that the maximum accuracy remains $1$ for $t\geq N$ in Fig.~\ref{fig:oneUpdate_empty_comparison_p}(d). The latter is because the new link must always involve the most accurate agent. We find that $c_t$ eventually decays to $0$ but after a long period of fluctuation. As an example, the numerical result for $p=0.1$ is presented in Fig.~\ref{fig:oneUpdate_empty_p0_1}(a--c). After $t=398$ ($\approx 2N$), $c_t$ begins to increase and then fluctuates before decaying to $0$. This fluctuating behavior determines the structure of stable link density as a function of $c$. We find two jumps in the stable link density, i.e., one at $c=c_0\approx 0.014$ and the other at $c=c_{398}\approx 0.00025$, corresponding to the deep valley of $c_t$ in $t=398$. It implies that the stationary network can be either empty for large $c$, a ring structure for intermediate $c$, or a complicated structure for small $c$. 

In general, the number of jumps in the stable link density may indicate the sensitivity on the cost of linking, hence the dynamical complexity of the network evolution. In Fig.~\ref{fig:oneUpdate_empty_p0_1}(d), we find that as $p$ increases, the number of jumps is increasing for $p<0.01$ and then decreasing for $p>0.01$. Here jumps are detected when the gap at the discontinuous transition is larger than $10^{-3}M$ corresponding to $20$ links. Since the smaller $p$ generally leads to more branches and loops in the early stage of dynamics, it is expected to result in the larger number of jumps. This explains only the decreasing behavior for $p>0.01$. To explain the increasing behavior for $p<0.01$, we note that as $p$ decreases, $c_0$ is also decreasing, see Fig.~\ref{fig:oneUpdate_empty_comparison_p}(c). If $c_0$ becomes small enough that $c_t$ for $t>0$ rarely decreases below $c_0$ before decaying to $0$, we find the smaller number of jumps in the stable link density, despite the large fluctuations in $c_t$. As $p$ approaches $0$, the average number of jumps reduces to $1$, meaning that the stationary network can be either empty for large $c$ or a complicated structure for small $c$. The non-monotonic behavior in the number of jumps according to $p$ may imply that there exists an optimal degree of perceptual overestimation of the initial link density to maximize the sensitivity on $c$. 


\begin{figure}[!t]
    \includegraphics[width=\columnwidth]{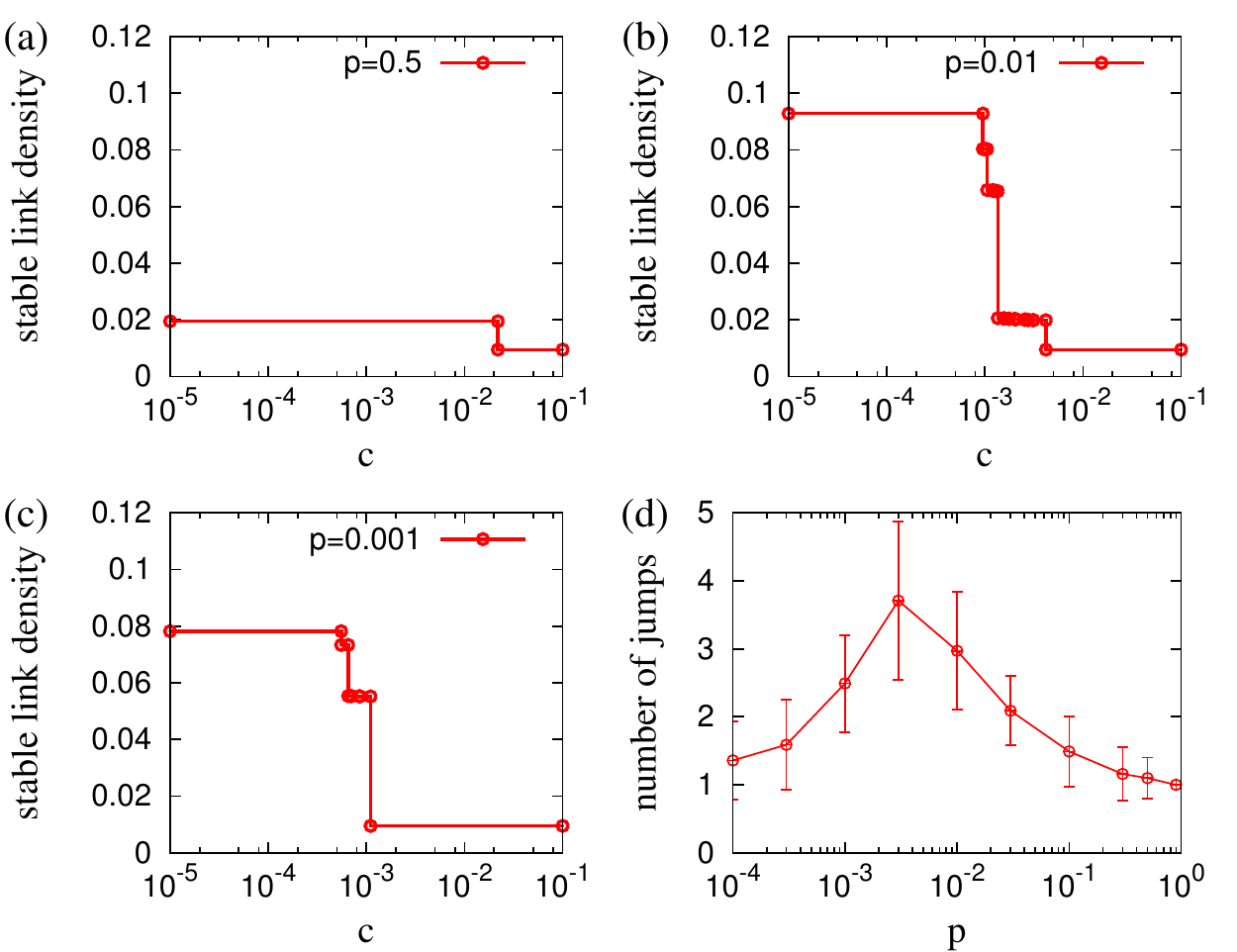}
    \caption{Single update case for an initially random network with linking probability $q=0.01$ for $N=200$ and for various values of $p$. In each panel of (a--c), we present the stable link density as a function of $c$. (d) The average number of jumps in the stable link density as a function of $c$ over $100$ runs is plotted with standard deviation for each value of $p$.}
    \label{fig:oneUpdate_ER0_01_comparison_p}
\end{figure}

Based on the case with $q=0$, we study the generalized case with $q>0$. For example, when $q=0.01$, each agent initially randomly knows a few of other agents. We find the qualitatively same results as in the case with $q=0$, except that the stationary network looks random since the actual network has been initially random without a ring structure. Irrespective of the initial structure of the actual network, the ring structure must be formed in the early stage of the dynamics if $c<c_0$. Thus, when $p$ is sufficiently large, the stationary network is either the initial random network for large $c$ or the initial random network with the ring for small $c$, see Fig.~\ref{fig:oneUpdate_ER0_01_comparison_p}(a) for the case with $p=0.5$. For smaller values of $p$, the stable link density as a function of $c$ shows the similar behavior as in the case with $q=0$, but elevated by $q$, as depicted in Fig.~\ref{fig:oneUpdate_ER0_01_comparison_p}(b,c). Accordingly, we find the non-monotonous behavior for the average number of jumps in the stable link density in Fig.~\ref{fig:oneUpdate_ER0_01_comparison_p}(d). We remark that our findings are robust with respect to the variation of $N$ and $q$.

\begin{figure}[!t]
  \includegraphics[width=\columnwidth]{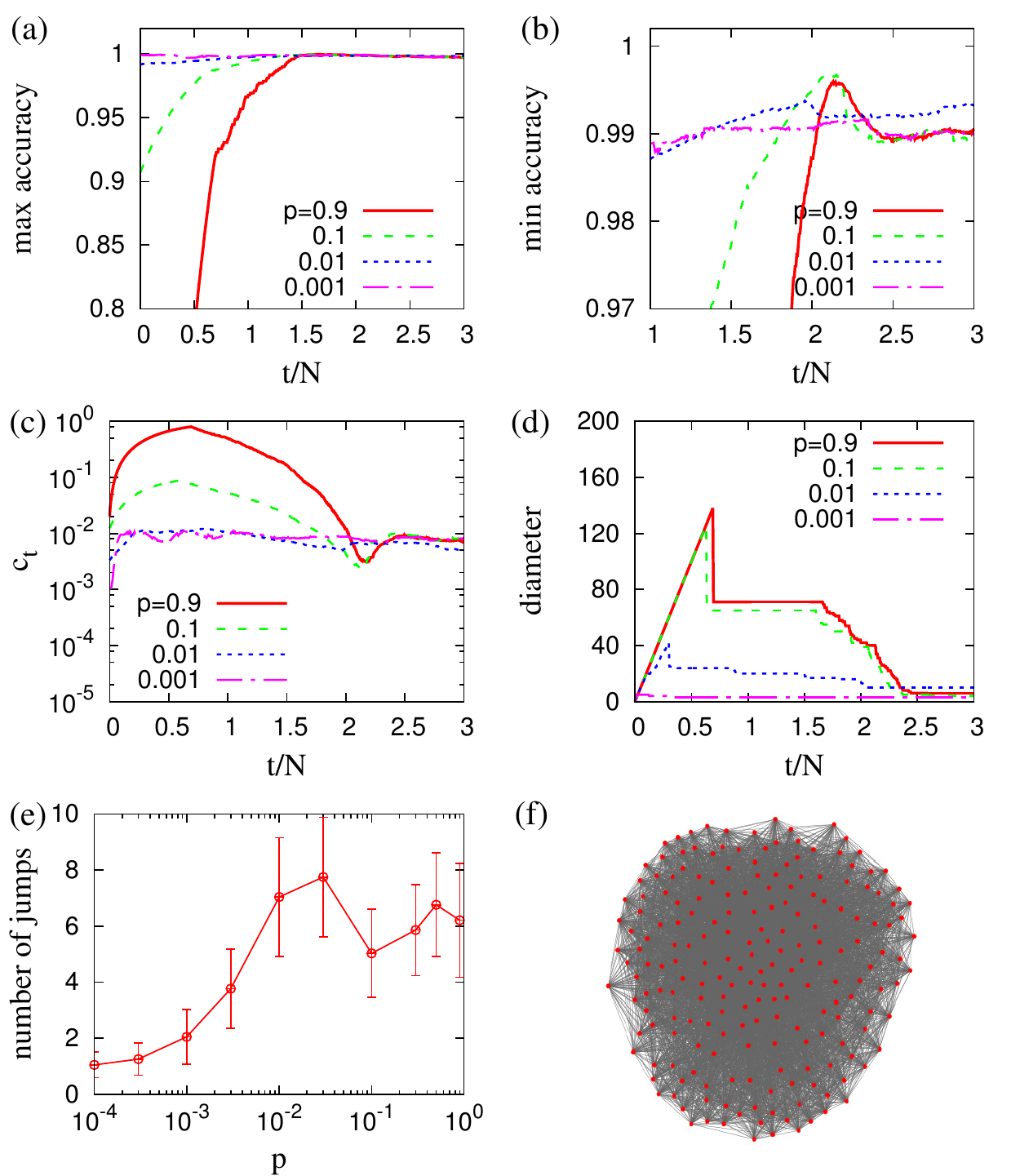}
  \caption{Multiple update case for an initially empty network for $N=200$ and for various values of $p$. We present temporal evolutions of (a) maximum accuracies, (b) minimum accuracies, (c) values of $c_t$ in Eq.~(\ref{eq:c_t}), and (d) diameters of actual networks. In (e), the average number of jumps in the stable link density as a function of $c$ over $100$ runs is plotted with standard deviation for each value of $p$. In (f), the stationary network for $p=0.1$ and $c=0$ is visualized.}
    \label{fig:multiUpdate_empty_comparison_p}
\end{figure}

\subsection{Multiple update case}

Next we consider the multiple update case in which more than one link can be created simultaneously as all agents in $L_t$ have the chance to update their perceptions using those of agents in $H_t$. 

When the actual network is initially empty, i.e., if $q=0$, due to a number of branches, the ring structure emerges considerably earlier than $t=N$ especially for small values of $p$, as depicted in Fig.~\ref{fig:multiUpdate_empty_comparison_p}(d). Figure~\ref{fig:multiUpdate_empty_comparison_p}(a,b) shows that the maximum accuracy approaches $1$ but may not reach $1$ for the entire range of $p>0$, which is also the case for the minimum accuracy around in $t=2N$. Thus, $c_t$ does not decay to $0$ around in $t=2N$, but fluctuates for a long time before eventually decaying to $0$. This behavior is observed even for $p$ that is very close to $1$, in contrast to the single update case. As a result, the number of jumps does not decrease but fluctuates for $p>0.01$, as shown in Fig.~\ref{fig:multiUpdate_empty_comparison_p}(e), while the increasing number of jumps for $p<0.01$ can be explained similarly to the single update case. In addition, in most cases, the link density of the stationary network for $c=0$ is high, e.g., it is $\approx 0.36$ for $p=0.1$. The stationary network for $p=0.1$ and $c=0$ is visualized in Fig.~\ref{fig:multiUpdate_empty_comparison_p}(f).

When the actual network is initially a random network with $q>0$, we find the qualitatively same behavior as in the case with $q=0$, except that the stationary network does not show a ring structure because initial networks are random. It would mean that the effect of multiple update rule dominates that of initial conditions controlled by $q$ and $p$.

\begin{figure}[!t]
  \includegraphics[width=\columnwidth]{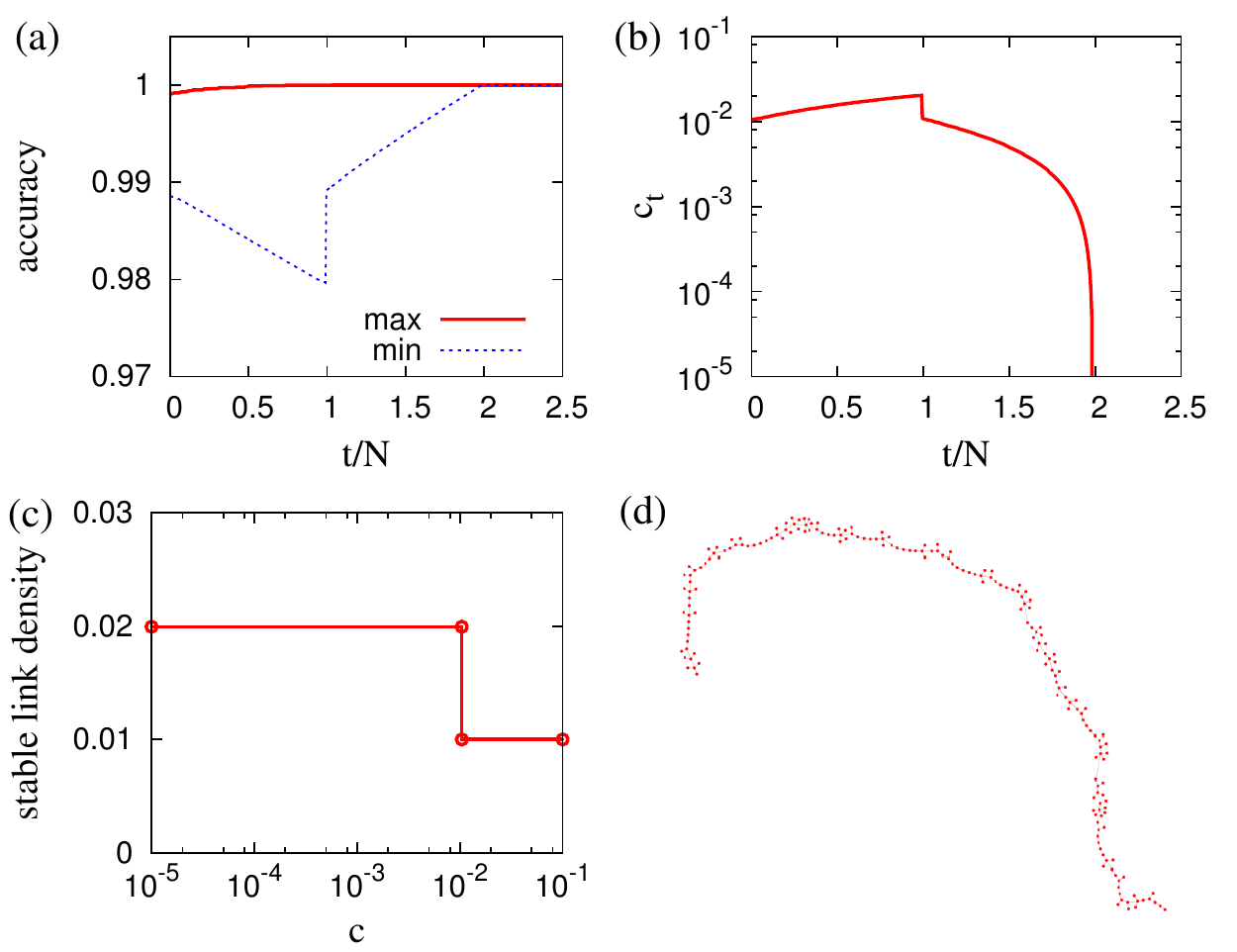}
  \caption{Single update case for an initially star network for $N=200$ and for $p=0.001$. We present temporal evolutions of (a) maximum and minimum accuracies, (b) the value of $c_t$ in Eq.~(\ref{eq:c_t}), with the resulting stable link density as a function of $c$ in (c). The stationary network for $c=0$ without the center agent is visualized in (d).}
    \label{fig:star_oneUpdate_empty_p0_001}
\end{figure}

\section{Heterogeneous initial condition: star networks}\label{sect:hetero}

We now consider the heterogeneous initial condition for the actual network, implying that only a few agents are highly connected, while most agents are poorly connected. For the simplest setup, we take a star network, where one agent, called center, is connected to all other agents. In general, the center is expected to be advantageous because it does not need to create a link when it is the least accurate agent. In addition, when the center is the most accurate agent, no new link is formed as the least accurate agent can use the existing link to the center for updating.

For the single update case, when $p$ is sufficiently large, the center is not necessarily the most accurate agent in $t=0$. Thus, the early stage dynamics is the same as in the case for the initially empty network, except for two time steps. In the time step when the center becomes the least accurate agent, the center updates its perception using that of the most accurate agent without creating a link. Note that the center does not yet have fully correct link states. In the next time step, the least accurate agent updates its perception using that of the center without creating a link. Therefore, we observe the qualitatively same behaviors for the maximum and minimum accuracies, hence for $c_t$ as in Fig.~\ref{fig:oneUpdate_empty_p0_5}(a,c). The stationary network turns out to have a wheel structure that is a line network connecting all agents on the top of the star network. Here the center is a part of the line network.

\begin{figure}[!t]
  \includegraphics[width=\columnwidth]{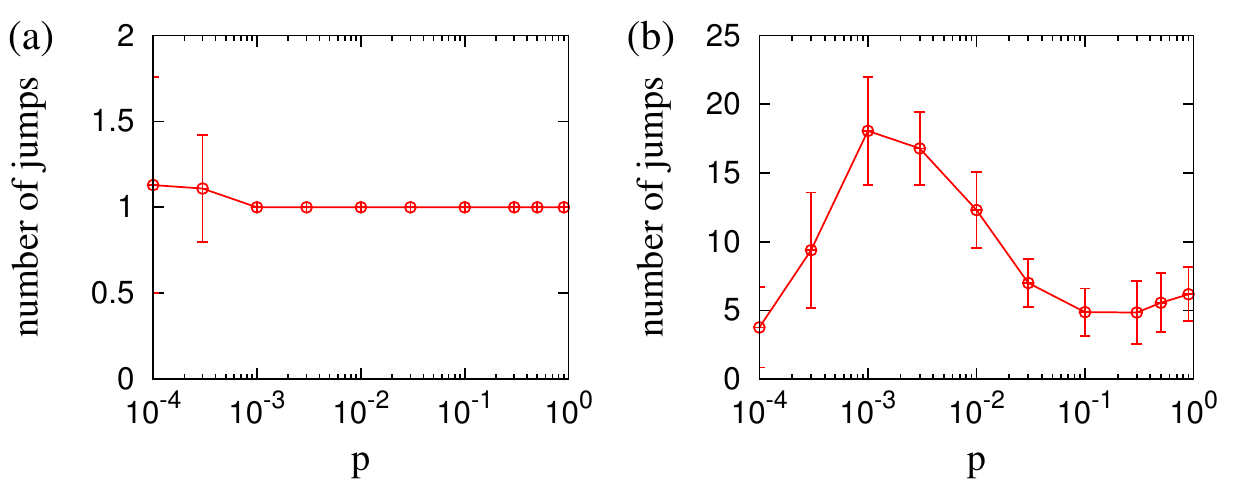}
  \caption{Single (a) and multiple (b) update cases for an initially star network for $N=200$. The average number of jumps of stable link density as a function of $c$ over $100$ runs is plotted with standard deviation for each value of $p$.}
    \label{fig:star_jump}
\end{figure}

When $p$ is sufficiently small, the center becomes the most accurate agent in $t=0$ as all other agents underestimate the existence of links involving the center. Once the least accurate agent updates its perception using that of the center, it is likely to become the most accurate agent in the next time step. Then, the dynamics is again similar to the case for the initially empty network but only for $t<N$. In $t\approx N$, the center becomes the most accurate agent with fully correct link states. Since then, the center remains as the most accurate agent for $N<t<2N$, while all other agents mostly update their perceptions using that of the center without creating any additional links. Thus, the stable link density remains $\approx \frac{4}{N}$. This behavior is clearly different from the case with homogeneous initial conditions. Figure~\ref{fig:star_oneUpdate_empty_p0_001} shows that when $p=0.001$, the temporal evolutions of maximum and minimum accuracies and of $c_t$ are consistent with the above argument, and that the stationary network has an overall wheel structure with many branches. Here the number of jumps is $1$. The average number of jumps remains $1$ for the almost entire range of $p$ as shown in Fig.~\ref{fig:star_jump}(a). It indicates that the sensitivity of the dynamics on $c$ is rarely affected by the degree of perceptual overestimation or underestimation of the initial link density, due to the crucial role of the center as an information spreader.

For the multiple update case, we find much more complex dynamics of $c_t$ than the previous cases for the almost entire range of $p$. This leads to overall larger numbers of jumps, hence to the higher sensitivity of the dynamics on $c$, as evidenced in Fig.~\ref{fig:star_jump}(b). For large values of $p$, the dynamics is similar to those in the multiple update case with homogeneous initial conditions, resulting in the similar numbers of jumps around $5$. This is due to the fact that the center no longer plays an information spreader as it cannot follow created links between other agents. Then, as $p$ decreases, the early stage dynamics becomes more complex to show the larger number of jumps. However, for sufficiently small $p$, we observe that $c_t$ has a relatively deep valley in the early stage of dynamics, implying that the minimum accuracy gets close to the maximum accuracy. It might be because the role of the center becomes crucial at least in the early stage of dynamics. Due to the deep valley in $c_t$ the number of jumps is considerably reduced. We remark that the smaller number of jumps for smaller $p$ in the homogeneous case is mainly due to the decreasing $c_0$ for small $p$, while in the heterogeneous case $c_0$ has a lower bound of $\frac{2(N-2)}{N(N-1)}\approx 0.01$ for $p=0$~\footnote{If $p=0$, the accuracy of the center is $\rho^{h_0}_0=1$, while all other agents have the same minimum accuracy of $\rho^{l_0}_0=1-\frac{N-2}{M}$. It is because each non-center agent has incorrect link states for $N-2$ links between the center and other $N-2$ agents.}.

\section{Conclusions}\label{sect:conclusion}

In order to understand the evolution of social networks in terms of perception-based strategic link formation, we have studied the perception-based network formation model. We assume that each individual has her own perception of the actual network, and uses it to decide whether to create a link to another individual. An individual with the least perception accuracy can benefit from updating her perception using that of the most accurate individual via a new link. This benefit is compared to the cost of linking $c$ for deciding whether to create the link. Once a new link is created, it affects the accuracies of other individuals, leading to further evolution of the actual network. As for initial actual networks, we consider homogeneous and heterogeneous cases. The homogeneous initial actual network is modeled by Erd\H{o}s-R\'enyi (ER) random networks, while we take a star network for the heterogeneous case. In any cases, an individual perception of the actual network is modeled by an ER random network with linking probability $p$. Here $p$ quantifies how much individuals initially overestimate or underestimate the link density of the actual network. Then the stable link density of the actual network is expected to decrease as $c$ increases. We find that the stable link density shows discontinuous transitions or jumps according to the cost of linking, where the number of jumps is the consequence of the dynamical complexity.

For the homogeneous initial condition, we first consider the single update case in which only one of the least accurate agents has a chance to update her perception. As a benchmark, when the initial actual network is empty, the stationary network is either empty for large $c$ or a ring structure for small $c$, for sufficiently large values of $p$. It is because agents with the maximum and minimum accuracies are respectively unique in each time step. As $p$ becomes smaller, it is more likely to find more than one agent having the maximum accuracy in the early stage of dynamics, leading to a number of branches along the ring. This leads to much longer and complex dynamics before eventually reaching the stationarity. Hence we find a more complicated structure of the stable link density as a function of $c$, but only for $p>0.01$. As $p$ decreases below $0.01$, the number of jumps is decreasing, implying the simpler dynamics. For $p$ approaching $0$, the stationary network is either empty for large $c$ or a complex structure for small $c$.  Then we expand the model to a general setting that the initial actual network is an Erd\H{o}s-R\'enyi random network with positive linking probability. The results are overall similar to those for initially empty networks, except that the stationary network does not show a ring structure because the initial network was random. 

The multiple update case implies that all the least accurate agents have the chance to update their perceptions. Then the branches and loops can be formed in the early stage of the dynamics, irrespective of $p$. Accordingly, we find that the average number of jumps for $p>0.01$ is fluctuating in contrast to the single update case, while it is decreasing as $p$ decreases below $0.01$. 

Next we have considered a star network as the heterogeneous initial actual network. It is found that for the single update case, the dynamics is simple due to the crucial role of the center agent as an information spreader. Once the center has fully correct link states, it spreads the correct information to all other agents without any new links. Thus the number of jumps remains $1$ irrespective of $p$. However, for the multiple update case, the role of the center strongly depends on the degree of perceptual estimation of the initial actual network, leading to a non-monotonic behavior of the dynamical complexity as a function of $p$.

In general, understanding human decision making in terms of perception is highly important, and our model can provide a useful framework for modeling perception of networks in physics and social sciences. Further extensions of our model can be considered to apply to relevant topics, e.g., modeling perception-based organizational behaviors in corporate governance, marketing strategies using the perception-based network, and explaining educational performances in our framework.

Finally, we remark that the aim of our work is to investigate the fundamental principles of human decision making by devising the simplest, possibly tractable, and yet nontrivial model, inevitably at the expense of the representation of the real social networks. We leave for future works how such fundamental principles of human decision making are translated into the empirical findings of social networks, as it must base on more comprehensive understanding of how people perceive the network, and how such perception affects link formation or deletion.

\begin{acknowledgments}
     H.-H.J. acknowledges financial support by Basic Science Research Program through the National Research Foundation of Korea (NRF) grant funded by the Ministry of Education (2015R1D1A1A01058958).
\end{acknowledgments}


\begin{thebibliography}{21}
\expandafter\ifx\csname natexlab\endcsname\relax\def\natexlab#1{#1}\fi
\expandafter\ifx\csname bibnamefont\endcsname\relax
  \def\bibnamefont#1{#1}\fi
\expandafter\ifx\csname bibfnamefont\endcsname\relax
  \def\bibfnamefont#1{#1}\fi
\expandafter\ifx\csname citenamefont\endcsname\relax
  \def\citenamefont#1{#1}\fi
\expandafter\ifx\csname url\endcsname\relax
  \def\url#1{\texttt{#1}}\fi
\expandafter\ifx\csname urlprefix\endcsname\relax\def\urlprefix{URL }\fi
\providecommand{\bibinfo}[2]{#2}
\providecommand{\eprint}[2][]{\url{#2}}

\bibitem[{\citenamefont{Albert and Barab\'{a}si}(2001)}]{Albert2001Statistical}
\bibinfo{author}{\bibfnamefont{R.}~\bibnamefont{Albert}} \bibnamefont{and}
  \bibinfo{author}{\bibfnamefont{A.-L.} \bibnamefont{Barab\'{a}si}},
  \bibinfo{journal}{Reviews of Modern Physics} \textbf{\bibinfo{volume}{74}},
  \bibinfo{pages}{47} (\bibinfo{year}{2001}).

\bibitem[{\citenamefont{Borgatti et~al.}(2009)\citenamefont{Borgatti, Mehra,
  Brass, and Labianca}}]{Borgatti2009Network}
\bibinfo{author}{\bibfnamefont{S.~P.} \bibnamefont{Borgatti}},
  \bibinfo{author}{\bibfnamefont{A.}~\bibnamefont{Mehra}},
  \bibinfo{author}{\bibfnamefont{D.~J.} \bibnamefont{Brass}}, \bibnamefont{and}
  \bibinfo{author}{\bibfnamefont{G.}~\bibnamefont{Labianca}},
  \bibinfo{journal}{Science} \textbf{\bibinfo{volume}{323}},
  \bibinfo{pages}{892} (\bibinfo{year}{2009}).

\bibitem[{\citenamefont{Goyal}(2009)}]{Goyal2009Connections}
\bibinfo{author}{\bibfnamefont{S.}~\bibnamefont{Goyal}},
  \emph{\bibinfo{title}{Connections: an introduction to the network economy}}
  (\bibinfo{publisher}{Princeton University Press}, \bibinfo{year}{2009}).

\bibitem[{\citenamefont{Murase et~al.}(2015)\citenamefont{Murase, Jo,
  T\"{o}r\"{o}k, Kert\'{e}sz, and Kaski}}]{Murase2015Modeling}
\bibinfo{author}{\bibfnamefont{Y.}~\bibnamefont{Murase}},
  \bibinfo{author}{\bibfnamefont{H.-H.} \bibnamefont{Jo}},
  \bibinfo{author}{\bibfnamefont{J.}~\bibnamefont{T\"{o}r\"{o}k}},
  \bibinfo{author}{\bibfnamefont{J.}~\bibnamefont{Kert\'{e}sz}},
  \bibnamefont{and} \bibinfo{author}{\bibfnamefont{K.}~\bibnamefont{Kaski}},
  \bibinfo{journal}{PLoS ONE} \textbf{\bibinfo{volume}{10}},
  \bibinfo{pages}{e0133005} (\bibinfo{year}{2015}).

\bibitem[{\citenamefont{Onnela et~al.}(2007{\natexlab{a}})\citenamefont{Onnela,
  Saram\"{a}ki, Hyv\"{o}nen, Szab\'{o}, Argollo~de Menezes, Kaski,
  Barab\'{a}si, and Kert\'{e}sz}}]{Onnela2007Analysis}
\bibinfo{author}{\bibfnamefont{J.-P.} \bibnamefont{Onnela}},
  \bibinfo{author}{\bibfnamefont{J.}~\bibnamefont{Saram\"{a}ki}},
  \bibinfo{author}{\bibfnamefont{J.}~\bibnamefont{Hyv\"{o}nen}},
  \bibinfo{author}{\bibfnamefont{G.}~\bibnamefont{Szab\'{o}}},
  \bibinfo{author}{\bibfnamefont{M.}~\bibnamefont{Argollo~de Menezes}},
  \bibinfo{author}{\bibfnamefont{K.}~\bibnamefont{Kaski}},
  \bibinfo{author}{\bibfnamefont{A.-L.} \bibnamefont{Barab\'{a}si}},
  \bibnamefont{and}
  \bibinfo{author}{\bibfnamefont{J.}~\bibnamefont{Kert\'{e}sz}},
  \bibinfo{journal}{New Journal of Physics} \textbf{\bibinfo{volume}{9}},
  \bibinfo{pages}{179} (\bibinfo{year}{2007}{\natexlab{a}}).

\bibitem[{\citenamefont{Onnela et~al.}(2007{\natexlab{b}})\citenamefont{Onnela,
  Saram\"{a}ki, Hyv\"{o}nen, Szab\'{o}, Lazer, Kaski, Kert\'{e}sz, and
  Barab\'{a}si}}]{Onnela2007Structure}
\bibinfo{author}{\bibfnamefont{J.~P.} \bibnamefont{Onnela}},
  \bibinfo{author}{\bibfnamefont{J.}~\bibnamefont{Saram\"{a}ki}},
  \bibinfo{author}{\bibfnamefont{J.}~\bibnamefont{Hyv\"{o}nen}},
  \bibinfo{author}{\bibfnamefont{G.}~\bibnamefont{Szab\'{o}}},
  \bibinfo{author}{\bibfnamefont{D.}~\bibnamefont{Lazer}},
  \bibinfo{author}{\bibfnamefont{K.}~\bibnamefont{Kaski}},
  \bibinfo{author}{\bibfnamefont{J.}~\bibnamefont{Kert\'{e}sz}},
  \bibnamefont{and} \bibinfo{author}{\bibfnamefont{A.~L.}
  \bibnamefont{Barab\'{a}si}}, \bibinfo{journal}{Proceedings of the National
  Academy of Sciences} \textbf{\bibinfo{volume}{104}}, \bibinfo{pages}{7332}
  (\bibinfo{year}{2007}{\natexlab{b}}).

\bibitem[{\citenamefont{Ahn et~al.}(2010)\citenamefont{Ahn, Bagrow, and
  Lehmann}}]{Ahn2010Link}
\bibinfo{author}{\bibfnamefont{Y.-Y.} \bibnamefont{Ahn}},
  \bibinfo{author}{\bibfnamefont{J.~P.} \bibnamefont{Bagrow}},
  \bibnamefont{and} \bibinfo{author}{\bibfnamefont{S.}~\bibnamefont{Lehmann}},
  \bibinfo{journal}{Nature} \textbf{\bibinfo{volume}{466}},
  \bibinfo{pages}{761} (\bibinfo{year}{2010}).

\bibitem[{\citenamefont{Fortunato}(2010)}]{Fortunato2010Community}
\bibinfo{author}{\bibfnamefont{S.}~\bibnamefont{Fortunato}},
  \bibinfo{journal}{Physics Reports} \textbf{\bibinfo{volume}{486}},
  \bibinfo{pages}{75} (\bibinfo{year}{2010}).

\bibitem[{\citenamefont{Newman}(2002)}]{Newman2002Assortative}
\bibinfo{author}{\bibfnamefont{M.~E.~J.} \bibnamefont{Newman}},
  \bibinfo{journal}{Physical Review Letters} \textbf{\bibinfo{volume}{89}},
  \bibinfo{pages}{208701} (\bibinfo{year}{2002}).

\bibitem[{\citenamefont{Barab\'{a}si and Albert}(1999)}]{Barabasi1999Emergence}
\bibinfo{author}{\bibfnamefont{A.-L.} \bibnamefont{Barab\'{a}si}}
  \bibnamefont{and} \bibinfo{author}{\bibfnamefont{R.}~\bibnamefont{Albert}},
  \bibinfo{journal}{Science} \textbf{\bibinfo{volume}{286}},
  \bibinfo{pages}{509} (\bibinfo{year}{1999}).

\bibitem[{\citenamefont{Jeong et~al.}(2003)\citenamefont{Jeong, N\'{e}da, and
  Barab\'{a}si}}]{Jeong2003Measuring}
\bibinfo{author}{\bibfnamefont{H.}~\bibnamefont{Jeong}},
  \bibinfo{author}{\bibfnamefont{Z.}~\bibnamefont{N\'{e}da}}, \bibnamefont{and}
  \bibinfo{author}{\bibfnamefont{A.~L.} \bibnamefont{Barab\'{a}si}},
  \bibinfo{journal}{Europhysics Letters (EPL)} \textbf{\bibinfo{volume}{61}},
  \bibinfo{pages}{567} (\bibinfo{year}{2003}).

\bibitem[{\citenamefont{Kossinets and Watts}(2006)}]{Kossinets2006Empirical}
\bibinfo{author}{\bibfnamefont{G.}~\bibnamefont{Kossinets}} \bibnamefont{and}
  \bibinfo{author}{\bibfnamefont{D.~J.} \bibnamefont{Watts}},
  \bibinfo{journal}{Science} \textbf{\bibinfo{volume}{311}},
  \bibinfo{pages}{88} (\bibinfo{year}{2006}).

\bibitem[{\citenamefont{Kumpula et~al.}(2007)\citenamefont{Kumpula, Onnela,
  Saram\"{a}ki, Kaski, and Kert\'{e}sz}}]{Kumpula2007Emergence}
\bibinfo{author}{\bibfnamefont{J.~M.} \bibnamefont{Kumpula}},
  \bibinfo{author}{\bibfnamefont{J.~P.} \bibnamefont{Onnela}},
  \bibinfo{author}{\bibfnamefont{J.}~\bibnamefont{Saram\"{a}ki}},
  \bibinfo{author}{\bibfnamefont{K.}~\bibnamefont{Kaski}}, \bibnamefont{and}
  \bibinfo{author}{\bibfnamefont{J.}~\bibnamefont{Kert\'{e}sz}},
  \bibinfo{journal}{Physical Review Letters} \textbf{\bibinfo{volume}{99}},
  \bibinfo{pages}{228701} (\bibinfo{year}{2007}).

\bibitem[{\citenamefont{Jo et~al.}(2011)\citenamefont{Jo, Pan, and
  Kaski}}]{Jo2011Emergence}
\bibinfo{author}{\bibfnamefont{H.-H.} \bibnamefont{Jo}},
  \bibinfo{author}{\bibfnamefont{R.~K.} \bibnamefont{Pan}}, \bibnamefont{and}
  \bibinfo{author}{\bibfnamefont{K.}~\bibnamefont{Kaski}},
  \bibinfo{journal}{PLoS ONE} \textbf{\bibinfo{volume}{6}},
  \bibinfo{pages}{e22687} (\bibinfo{year}{2011}).

\bibitem[{\citenamefont{Murase et~al.}(2014)\citenamefont{Murase,
  T\"{o}r\"{o}k, Jo, Kaski, and Kert\'{e}sz}}]{Murase2014Multilayer}
\bibinfo{author}{\bibfnamefont{Y.}~\bibnamefont{Murase}},
  \bibinfo{author}{\bibfnamefont{J.}~\bibnamefont{T\"{o}r\"{o}k}},
  \bibinfo{author}{\bibfnamefont{H.-H.} \bibnamefont{Jo}},
  \bibinfo{author}{\bibfnamefont{K.}~\bibnamefont{Kaski}}, \bibnamefont{and}
  \bibinfo{author}{\bibfnamefont{J.}~\bibnamefont{Kert\'{e}sz}},
  \bibinfo{journal}{Physical Review E} \textbf{\bibinfo{volume}{90}},
  \bibinfo{pages}{052810} (\bibinfo{year}{2014}).

\bibitem[{\citenamefont{Milgram}(1967)}]{Milgram1967Smallworld}
\bibinfo{author}{\bibfnamefont{S.}~\bibnamefont{Milgram}},
  \bibinfo{journal}{Psychology Today} \textbf{\bibinfo{volume}{1}},
  \bibinfo{pages}{60} (\bibinfo{year}{1967}).

\bibitem[{\citenamefont{Krackhardt}(1987)}]{Krackhardt1987Cognitive}
\bibinfo{author}{\bibfnamefont{D.}~\bibnamefont{Krackhardt}},
  \bibinfo{journal}{Social Networks} \textbf{\bibinfo{volume}{9}},
  \bibinfo{pages}{109} (\bibinfo{year}{1987}).

\bibitem[{\citenamefont{Castellano et~al.}(2009)\citenamefont{Castellano,
  Fortunato, and Loreto}}]{Castellano2009Statistical}
\bibinfo{author}{\bibfnamefont{C.}~\bibnamefont{Castellano}},
  \bibinfo{author}{\bibfnamefont{S.}~\bibnamefont{Fortunato}},
  \bibnamefont{and} \bibinfo{author}{\bibfnamefont{V.}~\bibnamefont{Loreto}},
  \bibinfo{journal}{Reviews of Modern Physics} \textbf{\bibinfo{volume}{81}},
  \bibinfo{pages}{591} (\bibinfo{year}{2009}).

\bibitem[{\citenamefont{Sen and Chakrabarti}(2014)}]{Sen2014Sociophysics}
\bibinfo{author}{\bibfnamefont{P.}~\bibnamefont{Sen}} \bibnamefont{and}
  \bibinfo{author}{\bibfnamefont{B.~K.} \bibnamefont{Chakrabarti}},
  \emph{\bibinfo{title}{Sociophysics: an introduction}}
  (\bibinfo{publisher}{Oxford University Press}, \bibinfo{year}{2014}).

\bibitem[{\citenamefont{Jo and Moon}(2015)}]{Jo2015Coevolution}
\bibinfo{author}{\bibfnamefont{H.-H.} \bibnamefont{Jo}} \bibnamefont{and}
  \bibinfo{author}{\bibfnamefont{E.}~\bibnamefont{Moon}},
  \emph{\bibinfo{title}{Coevolution of a network and perception}}
  (\bibinfo{year}{2015}), \eprint{1409.1436},
  \urlprefix\url{http://arxiv.org/abs/1409.1436}.

\bibitem[{\citenamefont{Gross and Blasius}(2008)}]{Gross2008Adaptive}
\bibinfo{author}{\bibfnamefont{T.}~\bibnamefont{Gross}} \bibnamefont{and}
  \bibinfo{author}{\bibfnamefont{B.}~\bibnamefont{Blasius}},
  \bibinfo{journal}{Journal of The Royal Society Interface}
  \textbf{\bibinfo{volume}{5}}, \bibinfo{pages}{259} (\bibinfo{year}{2008}).

\end{thebibliography}

\end{document}